\documentstyle[11pt]{article}
\begin{document}
\title{\Large \bf Integrability of the Heisenberg Chains with
                    Boundary Impurities and Their Bethe Ansatz}
 \author{{ Boyu Hou$^a$, Kangjie Shi$^{a,b}$, Ruihong  Yue$^{a,c}$, Shaoyou  Zhao$^a$}\\
 {\footnotesize $^a$ Institute of modern physics, Northwest University, P.O.Box 105,
        Xi'an 710069, P.R. China}\\
  {\footnotesize $^b$ CCAST (World Laboratory), P.O. Box 8730, Beijing 100080, China}\\
  {\footnotesize $^c$ Institute of  Theoretical Physics, Academia Sinica, Beijing 100080, China}}
\maketitle
\begin {abstract}
{\sl  In this paper, we show the integrability of spin-1/2 XXZ Heisenberg chain
with two arbitrary spin boundary Impurities. By using the fusion method, we
generalize it to  the spin-1 XXZ chain. Then the
eigenvalues of 
Hamiltonians of these models are obtained by the means of Bethe
ansatz method.}\\
    {\bf Pacs:} 75.20H, 12.40E, 75.10D
\end  {abstract}
 {\bf 1. $\;\;\;$  Introduction}
\\
 
    Recently, more and more papers have  focused on the Kondo problem. It is
well known that the spin dynamics of the Kondo problem is equivalent to the dynamics 
of the spin chain with magnetic impurities{\cite{NK}}. 
Although  magnetic impurities play  important role in the model,
they usually destroy the integrability of the system.
So how to maintain the integrability of the quantum impurity
system is an important problem. Fruitful
achievement has been
obtained based on the methods of the Bosonization and
renormalization 
technique, conformal field theory and exact
diagonalization in this field [2-5].

The quantum inverse scattering method and  
the Bethe ansatz technique have been powerful tools to study 
the integrable impurity problems in 1-dimensional physical
system, 
such as Wang {\sl et al} 's papers about the vertex model[6-8]
and Frahm {\sl et al} and Links {\sl et al}'s series papers about the
$t-j$
model with impurities[9-15]. 
The Heisenberg chain is an important model 
in the integrable system so many papers pay their attention to it.
 Andrei and Johannesson first considered the integrable
Heisenberg chain with impurities under periodic boundary
condition{\cite{AJ}}.
 Then Lee and Schlottmann generalized their results to arbitrary
 spin impurities\cite{LS,SM}. But they have to present some
unphysical terms in the
Hamiltonian to maintain its integrability, though those terms may
be irrelevant{\cite{W}}. To the open boundary condition problem,  Gaudin
considered the nonlinear Schr${\ddot o}$dinger
 model and
the spin-1/2 Heisenberg chain with simple open boundaries{\cite{G}},
 then Schulz and Alcaraz{\cite{S,AB}}{\sl et al.} generalized it to 
Hubbard and other models. Wang have discussed  the properties of
the impurities  with arbitrary spin
coupled to the spin-1/2 XXX chain \cite{W}. 
The spin-1/2 XXZ chain coupled with spin-1/2 
impurities  has also  been discussed in Ref.\cite{HP}.
In Ref.\cite{Zhao}, the integrability of the spin-1 XXX
chain with arbitrary spin impurities has been investigated.

  In this paper, we study the integrability of the open Heisenberg chain 
coupled with
 arbitrary spin impurities. We discuss the spin-1/2 XXZ chain in 
the first part of the present
paper. The spin-1 case is presented in the second part. A brief discussion
about our results is given in the last section.
\\
\\
{\bf 2. $\;\;\;$  The spin-1/2 XXZ chain}
\\
\\
\par

  The $R$-matrix of spin-1/2 XXZ Heisenberg chain
can be written as
\begin{equation} R(u)=\left ( \begin{array}{cccc}
                      \sin(u+\eta) & 0& 0&0\\
                      0&\sin u & \sin\eta &0\\
                       0&\sin\eta & \sin u &0\\
                       0&0&0&\sin(u+\eta)   \end{array} \right ).
\end{equation}             
This $R$-matrix is regular  and satisfies the unitarity condition
$$R(u)R(-u)=\sin(u+\eta)\sin(-u+\eta)= \rho(u).$$
If we suppose the first and the second space of the $R$-matrix are auxiliary and 
quantum space respectively, and this $R$-matrix, as an operator
matrix , can also be written as $L$-operator form     
\begin{equation}
L_n (u)=\sum_{j=1}^4w_j\sigma^j\otimes\sigma_n^j \;\;,
\end{equation}
where 
\begin{eqnarray*}
           && w_1=w_2=\frac{1}{2}\sin v,\\
           && w_4-w_3=\sin u,\\
           && w_4+w_3=\sin(u+\eta),
\end{eqnarray*}
$$ \begin{array}{llll}
  \sigma^1=\left( \begin{array}{lc} 0&1\\ 1&0\end{array} \right ),&
  \sigma^2=\left( \begin{array}{lc} 0&-i\\ i&0 \end{array} \right ),&
  \sigma^3=\left( \begin{array}{lc} 1&0\\ 0&-1\end{array} \right ),&
  \sigma^4=\left( \begin{array}{lc} 1&0\\ 0&1 \end{array}\right ).
\end{array}$$
The $L$-operator and $R$-matrix  satisfy the following Yang-Baxter relations (YBR)
\begin{eqnarray}
&&   R_{12}(u-v)\stackrel{1}{L}_{i}(u)\stackrel{2}{L}_{i}(v)
       =\stackrel{2}{L}_{i}(v)\stackrel{1}{L}_{i}(u)R_{12}(u-v),\nonumber\\ 
&&   R_{12}(u-v)\stackrel{1}{T}(u)\stackrel{2}{T}(v)
       =\stackrel{2}{T}(v)\stackrel{1}{T}(u)R_{12}(u-v),
\end{eqnarray}
where $R_{12}$ acts on the auxiliary space $v_1\otimes v_2$, and the
$T$ defined by $$T(u)=\prod _{j=1}^{N}L_{j}(u)$$ acts
on
quantum spaces as $v_1\otimes v_2\otimes\cdots\otimes v_N$. Here we 
have used the notation 
$\stackrel{1}{A}=A\otimes 1,$ $\stackrel{2}{A}=1\otimes A.$
  
In order to construct the open boundary condition consistent with the
integrability, we consider the reflection  equation
\begin{equation}
R_{12}(u-v)\stackrel{1}{K}(u)R_{21}(u+v)\stackrel{2}{K}(v)
         =\stackrel{2}{K}(v)R_{12}(u+v)\stackrel{1}{K}(u)R_{21}(u-v).
\end{equation}
 where $K$ is the reflecting matrix, which determines
the boundary terms in the Hamiltonian. One can prove that the double-row
monodromy
matrix defined by $U(u)=T(u)K(u)T^{-1}(-u)$ also satisfies the reflection equation
\begin{equation}
R_{12}(u-v)\stackrel{1}{U}(u)R_{21}(u+v)\stackrel{2}{U}(v)
          =\stackrel{2}{U}(v)R_{12}(u+v)\stackrel{1}{U}(u)R_{21}(u-v).
 \end{equation}

The dual $K$-matrix $K^+(u)$ can be defined by the automorphism{\cite{SK}}
 \begin{equation}
    \phi:  K(u)\rightarrow K^+(u)=K^t(-u-\eta).
 \end{equation}
It satisfies the dual reflection equation
 \begin{eqnarray}
&&R_{12}(-u+v)\stackrel{1}{K^+}(u)R_{12}(-u-v-2\eta)\stackrel{2}{K^+}(v)\nonumber\\
 && =\;\;\stackrel{2}{K^+}(v)R_{12}(-u-v-2\eta)\stackrel{1}{K^+}(u)R_{12}(-u+v).
\end{eqnarray}
Then the transfer matrix can be defined by
\begin{equation}
t(u)=trK^+(u)U(u).
\end{equation}
One can check that it satisfies the commutation relation
\begin{equation}[t(u),t(v)]=0.\end{equation}

 Now we couple the spin-1/2 XXZ chain with two arbitrary spin
impurities located at the ends of the system. Then the
$L$-operator of the boundary
cites can be written as
\begin{equation}                  
       L_i(u)=\left ( \begin{array}{cc}                  
              \sin(u+\frac{\eta}{2}+d_i^z\eta) & d_i^-\sin{\eta}\\
               d_i^+\sin{\eta} & \sin(u+\frac{\eta}{2}-d_i^z\eta)\end{array}\right ),
  (i=a,\;b)
\end{equation}
 where $d^{\pm}, d^z$ are components of an arbitrary spin $m$ of
$SU_q(2)$. 
One can easily check that the $L$-operator satisfies the first relation of (3).
It also have the unitarity relation
 $$ L_{i}(u)L_{i}(-u)
      =d^2\sin^2\eta+\sin(u+\frac{\eta}{2})\sin(-u+\frac{\eta}{2})=\rho _d(u) $$
with $d^2=\sin(m\eta)\sin(\eta+m\eta)/\sin^2\eta$.

Define
 \begin{eqnarray}
  T(u)&=&L_{b}(u+c_b)L_{N}(u)\cdots
      L_{2}(u)L_{1}(u)L_{a}(u+c_a),\nonumber\\
  \tilde{T}(u)&=&T^{-1}(-u)\times const. \nonumber\\
&=&L_{a}(u-c_a)L_{1}(u)L_{2}(u)\cdots L_{N}(u)L_{b}(u-c_b).
\end{eqnarray}  
where $c_i$ are free parameters. 
According to Cherednik\cite{Ch} and Sklyanin's 
work\cite{SK}, the reflection 
matrix and its dual are defined by  
\begin{equation}  
  K(u)=diag\left(1, \frac{\sin(\xi-u)}{\sin(\xi+u)}\right),
  K^+(u)=diag\left(1, \frac{\sin(\xi^++u+\eta)}{\sin(\xi^+-u-\eta)}\right).
\end{equation}
   Recalling the definition of $t(u)$ (8), one can
check that the above formulas satisfy the commutation relation (9).
By expanding $t(u)$ in terms of $u$, we can obtain infinite number of
conserved quantities which ensures the integrability of the model. The
Hamiltonian of this model can be written as
\begin{eqnarray}
H&=&\frac{1}{2\rho^{2N}(0)\rho_d(c_a)\rho_d(c_b)tr_\tau
    K^+(0)}\times\frac{d\;t(u)}{du}|_{u=0} \nonumber\\
&=&\sum_{j=1}^{N-1}\frac{H_{j,j+1}(u)}{\rho^{\frac{1}{2}}(u)}|_{u=0}\nonumber\\
& &+\;\frac{d(\stackrel{1}{L}_{a}(u+c_a)\stackrel{1}{L}(u-c_a))}
                                           {2\rho_{d}(c_a)du}|_{u=0}
    +\stackrel{1}{L}(u+c_a)\frac{dK_1(u)} 
           {2\rho_d(c_a)du}\stackrel{1}{L}(u-c_a)|_{u=0}\nonumber\\
& &+\;\frac{tr_{\tau}\stackrel{\tau}{K^+}(u)\stackrel{\tau}{L}_{b}(u+c_b)
                    H_{\tau,N}(u)\stackrel{\tau}{L}_{b}(u-c_b)}
   {\rho_d(c_b)\rho^{\frac{1}{2}}(u)tr_{\tau}\stackrel{\tau}{K^+}(u)}|_{u=0}\nonumber\\
& &+\;\frac{tr_{\tau}\stackrel{\tau}{K^+}(u)
            d(\stackrel{\tau}{L}_{b}(u+c_b)\stackrel{\tau}{L}_{b}(u-c_b))}
                {2\rho_d(c_b)tr_{\tau}\stackrel{\tau}{K^+}(u) du}
             |_{u=0}+const.
\end{eqnarray}
where $H_{j,j+1}(u)=\frac{dR_{j,j+1}(u)}{du}R_{j,j+1}(u).$ 
Denoting by  $T_i$  the $i$th term of the right
hand side of (13), we have
\begin{eqnarray}
&&T_1=\frac{1}{\sin\eta}\sum_{j=1}^{N-1}\left(\sigma ^1_j\cdot\sigma
               ^1_{j+1}+\sigma^2_j\cdot\sigma
       ^2_{j+1}+\cos\eta\sigma^3_j\cdot\sigma^3_{j+1}
\right), \nonumber\\
&&T_2+T_3=\frac{1}{2}(1+\sigma^3_1)\cdot A_a
          +\frac{1}{2}(1-\sigma_1^3)\cdot B_a
          +\sigma_1^+\cdot C_a d_a^-+\sigma_1^-\cdot d_a^+
C_a,\nonumber\\
&&T_4+T_5=\frac{1}{2}(1+\sigma_N^3)\cdot A_b
         +\frac{1}{2}(1-\sigma_N^3)\cdot B_b
          +\sigma_N^+\cdot C_b d_b^-+\sigma_N^-\cdot d_b^+C_b,
             \nonumber\\
&&A_i=\frac{1}{2\rho_d(c_i)}\left (\sin(\eta+2d^z\eta)-\frac{2\cos\xi}{\sin\xi}
 (d^2\sin^2(\eta)-\sin(d^z\eta)\sin(\eta+d^z\eta))
\right ),
                  \nonumber\\
&&B_i=\frac{1}{2\rho_d(c_i)}\left (\sin(\eta-2d^z\eta)-\frac{2\cos\xi}{\sin\xi}
\sin(c_i+\frac{\eta}{2}-d^z\eta)\sin(-c_i+\frac{\eta}{2}-d^z\eta)
           \right),\nonumber\\
&&C_i=-\frac{2\cos\xi}{\rho_d(c_i)\sin\xi}
             \sin(c_i+\frac{\eta}{2}-\xi+d^z\eta),\;\; (i=a,\;b)
\end{eqnarray}
where $\xi$ should be changed to $\xi^+$ when $i=b$, and this Hamiltonian is
hermitician
when we choose pure imaginary $c_i$.  

    To construct the algebraic Bethe ansatz, we rewrite the double-row monodromy 
matrix $U(u)$ in the form
\begin{equation}
  U(u)=\left ( \begin{array}{cc}
                     {\cal A}(u)&{\cal B}(u)\\ {\cal C}(u)&{\cal D}(u)\end{array}
\right ).\end{equation}
 Using the reflection equation (5), we can obtain the following commutation relation
\begin{eqnarray} 
{\cal B}(u){\cal B}(v)&=&{\cal B}(v){\cal B}(u); \nonumber \\
{\cal A}(u){\cal B}(v)&=&\frac{\sin(\eta +v-u)\sin(u+v)}{\sin(\eta +v+u)\sin(v-u)}
                         {\cal B}(v){\cal A}(u) \nonumber \\
                      & &+\;\;\frac{\sin\eta\sin(u+v)}{\sin(\eta +v+u)\sin(u-v)}
                         {\cal B}(u){\cal A}(v) \nonumber \\
                      & &-\;\;\frac{\sin\eta}{\sin(\eta +v+u)}
                         {\cal B}(u){\cal D}(v); \nonumber \\
{\tilde{\cal D}}(u){\cal B}(v)&=&\frac{\sin(u-v+\eta)\sin(u+v+2\eta)}{\sin(u+v+\eta)\sin(u-v)}
                         {\cal B}(v){\tilde{\cal D}}(u) \nonumber \\
                      & &-\;\;\frac{\sin\eta\sin(2u+2\eta)}{\sin(u-v)\sin(2v+\eta)}
                         {\cal B}(u){\tilde{\cal D}}(v) \nonumber \\
                      & &+\;\;\frac{\sin\eta\sin(2v)\sin(2u+2\eta)}{\sin(2v+\eta)\sin(u+v+\eta)}
                         {\cal B}(u){\cal A}(v),
\end{eqnarray}     
where ${\tilde{\cal D}}(u)=\sin(2u+\eta){\cal D}(u)-\sin\eta{\cal A}(u).$
Using the relation
 (8) and (15), the transfer matrix t(u) can now be written as
\begin{equation}
t(u)=w_1^+{\tilde{\cal D}}(u)+w_2^+{\cal A}(u)
    =\frac{\sin(\xi^++u+\eta)}{\sin(2u+\eta)}{\tilde{\cal D}}(u)
    +\frac{\sin(\xi^+-u)\sin(2u+2\eta)}{\sin(2u+\eta)}{\cal A}(u).
\end{equation}
Define the pseudo-vacuum state $\vert 0\rangle$
 \begin{equation}
     \sigma_i ^+\vert 0\rangle=d^+\vert 0\rangle=0,
                  \;\;(i=1,\;\;2,\;\;\cdots,\;\;N) 
 \end{equation}
Acting the elements of $U_{\tau}(u)$ on $\vert 0\rangle$, we have 
\begin{eqnarray}
 {\cal C}(u)\vert 0\rangle&=&0 \nonumber \\
 {\cal A}(u)\vert 0\rangle&=&w_1\vert 0\rangle \nonumber \\
       &=&\sin(\xi+u)\sin^{2N}(u+\eta)\prod_{r=\pm 1}
              \prod_{i=a,b}\sin(u+rc_i+\eta/2+m\eta)
                  \vert 0\rangle \nonumber \\
 {\tilde{\cal D}}(u)\vert 0\rangle&=&w_2\vert 0\rangle\nonumber \\
       &=&\sin(\xi-u-\eta)\sin(2u)\sin^{2N} u \prod_{r=\pm
1}\prod_{i=a,b}\sin(u+rc_i+\eta/2-m\eta)
                   \vert 0\rangle
\end{eqnarray}
The eigenstates of $t(u)$ can be constructed from the pseudo-vacuum state 
 \begin{equation}
  \vert\Omega\rangle =\prod_{i=1}^M{\cal B}(u_i)\vert 0\rangle.
 \end{equation}
Thus we obtain the eigenvalue of $t(u)$ acting on $\vert \Omega\rangle$
\begin{eqnarray}
 t(u)\vert\Omega\rangle&=&w_1w_1^+\prod_{i=1}^M\frac{\sin(\eta +v_i-u)\sin(u+v_i)}
                     {\sin(\eta+v_i+u)\sin(-u+v_i)}\vert \Omega\rangle \nonumber \\
                   &+&w_2w_2^+\prod_{i=1}^M\frac{\sin(u-v_i+\eta)\sin(u+v_i+2\eta)}
                     {\sin(u+v_i+\eta)\sin(u-v_i)}\vert \Omega\rangle 
\end{eqnarray}     
with the Bethe ansatz
\begin{eqnarray}
& &\frac{\sin(v_j-\xi^+-\frac{\eta}{2})\sin(v_j+\xi-\frac{\eta}{2})
           \sin^{2N}(v_j+\frac{\eta}{2})}
       {\sin(v_j+\xi^++\frac{\eta}{2})\sin(v_j-\xi+\frac{\eta}{2})\sin^{2N}
                     (v_j-\frac{\eta}{2})}\nonumber\\
&\times &\prod_{r=\pm 1}\prod_{k=a,b}\frac{\sin(v_j+rc_k+m\eta)}
                                 {\sin(v_j+rc_k-m\eta)}
=\prod_{i\ne j}^M\frac{\sin(v_j-v_i+\eta)\sin(v_j+v_i+\eta)}
                          {\sin(v_j-v_i-\eta)\sin(v_j+v_i-\eta)}.
\end{eqnarray}
\\
{\bf 3 $\;\;\;$ The spin-1 Heisenberg chain}
\\
{\bf 3.1 $\;\;\;$ The Zamolodchikov-Fateev 19-vertex model}
\\

 The Zamolodchikov-Fateev 19-vertex $R$-matrix\cite{ZF} associated with the
spin-1 representation of $U_q(sl_2)$\cite {Ku} can be obtained by 
using the fusion method\cite{ZF,Y}. It reads
\begin{equation}
R(u)= \left (
\begin{array}{ccccccccc}
a_1 & 0 & 0 & 0 & 0 & 0 & 0 & 0 & 0\\
0 & a_2 & 0 & a_3 & 0 & 0 & 0 & 0 & 0\\
0 & 0 & a_4 & 0 &a_5 & 0 & a_6 & 0 & 0\\
0 & a_3 & 0 & a_2 & 0 & 0 & 0 & 0 & 0\\
0 & 0 & a_5 & 0 & a_7 & 0 & a_5 & 0 & 0\\
0 & 0 & 0 & 0 & 0 & a_2 & 0 & a_3 & 0\\
0 & 0 & a_6 & 0 & a_5 & 0 & a_4 & 0 & 0\\
0 & 0 & 0 & 0 & 0 & a_3 & 0 & a_2 & 0\\
0 & 0 & 0 & 0 & 0 & 0 & 0 & 0 & a_1
\end{array} \right ),
\end{equation}
with $$\begin{array}{ll}
a_1=\sin(u+2\eta)\sin(u+\eta), & a_2=\sin u\sin(u+\eta),\\ 
a_3=\sin(2\eta)\sin(u+\eta),   & a_4=\sin u\sin(u-\eta), \\       
a_5=\sin u \sin(2\eta),        & a_6=\sin\eta\sin(2\eta),\\
a_7=a_2+a_6. \end{array}$$
This $R$-matrix is regular and satisfies the unitarity relation
$$\sin(u+\eta)\sin(u-\eta)\sin(u+2\eta)\sin(u-2\eta)=\rho(u). $$
It satisfies the Yang Baxer equation 
\begin{equation}
R_{12}(u-v)R_{13}(u)R_{23}(v)=R_{23}(v)R_{13}(u)R_{12}(u-v).
\end{equation}
 Inami {\sl et
al}\cite{ZH} have obtained the general solution $K(u)$ of (4). In this paper,
we only adopt its diagonal form as in the Ref.\cite{Mez}. 
\begin{eqnarray}
 K(u)&\equiv& diag(k_1(u),k_2(u),k_3(u)) \nonumber \\
     &=&diag\left( 1,\frac{\sin(\frac{\eta}{2}+\xi -u)}{\sin(\frac{\eta}{2}+\xi +u)},
                     \frac{\sin(\frac{\eta}{2}+\xi -u)\sin(\frac{\eta}{2}-\xi +u)}
                          {\sin(\frac{\eta}{2}+\xi +u)\sin(\frac{\eta}{2}-\xi -u)}\right
),
\end{eqnarray}
and the corresponding dual reflection matrix takes the form
\begin{eqnarray}
 K^+(u)&\equiv& diag(k_1^+(u),k^+_2(u),k^+_3(u)) \nonumber \\
     &=&diag\left( 1,\frac{\sin(\frac{3\eta }{2}+\xi^++u)}{\sin(-\frac{\eta
}{2}+\xi^+-u)},
                     \frac{\sin(\frac{3\eta}{2}+\xi^++u)\sin(-\frac{\eta
}{2}-\xi^+-u)}
{\sin(-\frac{\eta}{2}+\xi^+-u)\sin(\frac{3\eta}{2}-\xi^++u)}\right
).
\end{eqnarray}
\\
\\
{\bf 3.2 $\;\;\;$ Fusion of the boundary $L$-operator}
\\
\\

In this section we discuss the fusion procedure of the boundary $L$-operator. The
permutation operator and the projection operators are  defined by 
\begin{eqnarray}
& &P_{12}=R_{12}(0)/\sin(\eta)\\
& &P_{12}^- = -R_{12}(-\eta)/(2\sin\eta) \\
& &P_{12}^+ = 1-P_{12}^-
\end{eqnarray}
 respectively, where the $R(u)$ is from the 6-vertex model (1). They satisfy the
following
properties
\begin{eqnarray}
&&P_{12}^2=1,\;\;\;\;(P_{12}^{\pm})^2=P_{12}^{\pm}, \nonumber\\
&& P_{12}^+P_{12}^-=P_{12}^-P_{12}^+=0. 
\end{eqnarray}

 Now we use fusion procedure to obtain the high-dimensional $L$-operator.  
Taking $v=u+\eta$ in the equation (3),the YBR gives
\begin{equation}
 R_{12}(-\eta)L_{1d}(u)L_{2d}(u+\eta)=L_{2d}(u+\eta)L_{1d}(u)R_{12}(-\eta),
\end{equation}
where d represent the boundary terms, and we write $\stackrel{i}{L}_j$ as
$L_{ij}$ for convenience.  Multiplying above equation by $P_{12}^+$ from 
the left and right respectively, we get
\begin{eqnarray}
& &P_{12}^+L_{2d}(u+\eta)L_{1d}(u)P_{12}^- =0,\\
& &P_{12}^-L_{1d}(u)L_{2d}(u+\eta)P_{12}^+ =0.
\end{eqnarray}
Define \begin{eqnarray}
&&L_{<12>d}(u)=P_{12}^+L_{1d}(u)L_{2d}(u+\eta)P_{12}^+,\\
&&L'_{<12>d}(u)=P_{12}^+L_{2d}(u)L_{1d}(u-\eta)P_{12}^+,
\end{eqnarray}
which satisfy the YBR respectively.
\begin{equation}
R_{<12><34>}(u-v)L_{<12>d}(u)L_{<34>d}(v)
   =L_{<34>d}(v)L_{<12>d}(u)R_{<12><34>}(u-v),\end{equation}
\begin{equation}
R'_{<12><34>}(u-v)L'_{<12>d}(u)L'_{<34>d}(v)  
    =L'_{<34>d}(v)L'_{<12>d}(u)R'_{<12><34>}(u-v),
\end{equation}
where $R_{<12><34>}(u-v)$ is the fused $R$-matrix\cite{Y}, acts on $V_{<12>}\otimes
V_{<34>}$.
Here we give the proof for R(u):
\begin{eqnarray}
LHS&=&R_{<12><34>}(u-v)P_{12}^+L_{1d}(u)L_{2d}(u+\eta)P_{12}^+
                       P_{34}^+L_{3d}(v)L_{4d}(v+\eta)P_{34}^+ \nonumber \\
&=&P_{12}^+P_{34}^+R_{14}(u-v-\eta)R_{13}(u-v)P_{34}^+P_{34}^+
 R_{24}(u-v)R_{23}(u-v+\eta) \nonumber \\
& &     \times P_{34}^+P_{12}^+
   L_{1d}(u)L_{2d}(u+\eta)L_{3d}(v)L_{4d}(v+\eta)P_{12}^+P_{34}^+\nonumber\\
&=&R_{14}(u-v-\eta)R_{13}(u-v)R_{24}(u-v)R_{23}(u-v+\eta)
           L_{1d}(u)L_{2d}(u+\eta)\nonumber\\
& &   \times L_{3d}(v)L_{4d}(v+\eta)P_{12}^+P_{34}^+ \nonumber\\
&=&R_{14}(u-v-\eta)R_{24}(u-v)R_{13}(u-v)L_{1d}(u)L_{3d}(v) L_{2d}(u+\eta)\nonumber\\
& &\times R_{23}(u-v+\eta)L_{4d}(v+\eta)P_{12}^+P_{34}^+ \nonumber\\
&=&L_{3d}(v)L_{4d}(v+\eta)L_{1d}(u)L_{2d}(u+\eta)R_{14}(u-v-\eta)R_{13}(u-v)\nonumber\\
& &\times R_{24}(u-v)R_{23}(u-v+\eta)P_{34}^+P_{12}^+\nonumber \\
&=&
P_{34}^+L_{3d}(v)L_{4d}(v+\eta)P_{34}^+P_{12}^+L_{1d}(u)L_{2d}(u+\eta)P_{12}^+
P_{12}^+P_{34}^+R_{14}(u-v-\eta)\nonumber \\
& &     \times R_{13}(u-v)P_{34}^+P_{34}^+
             R_{24}(u-v)R_{23}(u-v+\eta)P_{34}^+P_{12}^+\nonumber \\
&=& L_{<34>d}(v)L_{<12>d}(u)R_{<12><34>}(u-v)\nonumber \\
&=& RHS.
\end{eqnarray}                   
The proof for the other formula is similar. Substituting relation (10) into
(33) and taking the transformation 
\begin{equation}
L_{<12>d}(u):\rightarrow (1\;\;\sqrt{2\cos\eta}\;\;1)L_{<12>d}(u)\left (
     \begin{array}{c}1\\ \frac{1}{\sqrt{2\cos\eta}}\\ 1 \end{array}\right ),
\end{equation}
 we have
\begin{equation}
L_{<12>d}(u)=\left ( \begin{array}{ccc}
   a_{11}&a_{12}&a_{13}\\a_{21}&a_{22}&a_{23}\\ a_{31}&a_{32}&a_{33}
\end{array}\right ),
\end{equation}
with
\begin{eqnarray*}
a_{11}&=&\sin(u+d^z\eta)\sin(u+\eta+d^z\eta),\\
a_{12}&=&d^-\sqrt{2\cos\eta}\sin(u+d^z\eta)\sin\eta,\\
a_{13}&=&(d^-)^2\sin^2\eta,\\
a_{21}&=&\sqrt{2\cos\eta}\sin(u+d^z\eta)\sin\eta d^+,\\
a_{22}&=&\sin u\sin(u+\eta)+d^2\sin^2\eta-2\sin^2(d^z\eta)\cos\eta,\\
a_{23}&=&\sqrt{2\cos\eta}\sin(u-d^z\eta)\sin\eta d^-,\\
a_{31}&=&(d^+)^2\sin^2\eta,\\
a_{32}&=&d^+\sqrt{2\cos\eta}\sin(u-d^z\eta)\sin\eta,\\
a_{33}&=&\sin(u-d^z\eta)\sin(u+\eta-d^z\eta).
\end{eqnarray*}
This $L$-operator also satisfies the unitarity relation with 
\begin{eqnarray*} 
\rho_d(u)&=&\frac{1}{4}\{(-d^2-\cos(2u-\eta)+\cos\eta+d^2\cos(2\eta)\}\nonumber \\
         & &\;\;\times \{-d^2-\cos(2u+\eta)+\cos\eta+d^2\cos(2\eta) \}.
\end{eqnarray*}
 \\
{\bf 3.3 $\;\;\;$ The Hamiltonian of this model}
\\

Define 
 \begin{eqnarray}
  T(u)&=&L_{b}(u+c_b)L_{N}(u)\cdots
L_{2}(u)L_{1}(u)L_{a}(u+c_a),\nonumber\\
  \tilde{T}(u)&=&T^{-1}(-u)\times const.\\
       &=&L_{a}(u-c_a)L_{1}(u)L_{2}(u)\cdots L_{N}(u)L_{b}(u-c_b),
\end{eqnarray}
 where $c_a$ and $c_b$ are constant. The spin-1 $L$-operator is obtained
from the $R$-matrix (23) by
assigning the second space to be the quantum space, and $L_i(u)$
$(i=a,\;b)$ is given by (40). The Hamiltonian of this
model is as same as the spin-1/2 case (13). Here we give $T_i$ as  
\begin{eqnarray*} 
T_1&=&\frac{1}{\sin(2\eta)}\sum_{j=1}^{N-1}\left\{ \frac{\vec s_j\cdot\vec
s_{j+1}}{\cos\eta}-\frac{(\vec s_j\cdot\vec s_{j+1})^2}{\cos^2\eta}
+(1-\cos\eta)\left(s^z_js^z_{j+1}s_j^+s_{j+1}^-+s_j^-s_{j+1}^+\right)\right.
                 \nonumber\\
& &-\;\left.\left(1-\cos(2\eta)\right)\left(s^z_js^z_{j+1}
           -(s^z_j)^2(s^z_{j+1})^2+(s^z_j)^2+(s^z_{j+1})^2\right)\right\}
\end{eqnarray*}
with $$\frac{1}{\cos\eta}\vec  s_j\cdot \vec s_{j+1}=
\frac{1}{2}s_j^-s_{j+1}^++\frac{1}{2}s_j^-s_{j+1}^-
+\cos\eta\frac{\sin(s^z_j\eta)\sin(s^z_{j+1}\eta)}{\sin^2\eta},$$
and $$\begin{array}{lll}
s^+=\sqrt{2\cos\eta}\left ( \begin{array}{lll}
      0&1&0\\0&0&1\\0&0&0 \end{array}\right ),   &
s^-=\sqrt{2\cos\eta}\left ( \begin{array}{lll}
      0&0&0\\1&0&0\\0&1&0 \end{array}\right ),   &
s^z=\left ( \begin{array}{llc}
      1&0&0\\0&0&0\\0&0&-1 \end{array}\right ),   
\end{array}$$
\begin{eqnarray*}
T_2+T_3&=&\frac{1}{2}(s_z+s_z^2)A_a +\sin\eta s_zs^+D_ad^- +(s^+)^2E_a(d^-)^2\\
& & +\sin\eta s^-s_zd^+D_a
         +(\frac{1}{2\cos\eta}(s^+s^-+s^-s^+)-s_z^2)B_a -\sin\eta s^+s_zF_ad^-\\ 
& &+(s^-)^2(d^+)^2E_a-\sin\eta s_zs^-d^+F_a+\frac{1}{2}(s_z^2-s_z)C_a,
\end{eqnarray*}
\begin{eqnarray*}
T_4+T_5&=&\frac{1}{2}(s_z+s_z^2)A_b +\sin\eta s_zs^+D_bd^- +(s^+)^2E_b(d^-)^2\\
& & +\sin\eta s^-s_zd^+D_b
         +(\frac{1}{2\cos\eta}(s^+s^-+s^-s^+)-s_z^2)B_b -\sin\eta s^+s_zF_bd^-\\
& &+(s^-)^2(d^+)^2E_b-\sin\eta s_zs^-d^+F_b+\frac{1}{2}(s_z^2-s_z)C_b,
\end{eqnarray*}
with 
\begin{eqnarray*}
A_i&=&\frac{1}{2\rho_d(c_i)}
\left\{\cos(2c_i)\cos\eta\sin(\eta+2d^z\eta)-\frac{1}{2}\sin(2\eta+4d^z\eta)\right.\\
& &+[d^2\sin^2\eta-\sin(d^z\eta)\sin(\eta+d^z\eta)]\cdot 
                             [2\cos\eta\sin(2\eta+2d^z\eta)\\  
& &-\;\frac{4\cos(\frac{\eta}{2}+\xi)
                    \cos\eta\sin(c_i+\eta+d^z\eta)\sin(-c_i+\eta+d^z\eta)}
           {\sin(\frac{\eta}{2}+\xi)}\\
& &\left. +\;\frac{2\sin(2\xi)(d^2\sin^2\eta-\sin(\eta+d^z\eta)\sin(2\eta+d^z\eta))}
        {\sin(\frac{\eta}{2}-\xi)\sin(\frac{\eta}{2}+\xi)}]\right \},\\
B_i&=&\frac{1}{2\rho_d(c_i)}
      \left \{2\sin\eta \sin(2c_i)[d^2\sin^2\eta-\cos\eta\sin(d^z\eta)]
               +\sin(2\eta)(\sin^2(2d^z\eta)-\sin^2c_i) \right. \\
& &-\;\frac{2\cos(\frac{\eta}{2}+\xi)}{\sin(\frac{\eta}{2}+\xi)}
     \times(d^2\sin^2\eta+\sin c_i\sin(c_i+\eta)-2\cos\eta\sin^2(d^z\eta)\\
& &\times (d^2\sin^2\eta+\sin c_i\sin(c_i-\eta)-2\cos\eta\sin^2(d^z\eta)\\
& &\left. -\;\frac{4\sin(2\xi)\cos\eta\sin(c_i-d^z\eta)\sin(-c_i-d^z\eta)
(d^2\sin^2\eta-\sin(d^z\eta)\sin(\eta+d^z\eta))}
   {\sin(\frac{\eta}{2}-\xi)\sin(\frac{\eta}{2}+\xi)}\right
\},\\
C_i&=&\frac{1}{2\rho_d(c_i)}
   \left \{\cos(2c_i)\cos\eta\sin(\eta-2d^z\eta)-\frac{1}{2}\sin(2\eta-4d^z\eta)
                                                 \right. \\
& &+[d^2\sin^2\eta+\sin(d^z\eta)\sin(\eta-d^z\eta)]
                        \cdot [2\cos\eta\sin(2\eta-2d^z\eta)\\ 
& &+\;\frac{4\cos(\frac{\eta}{2}+\xi)
                        \cos\eta\sin(c_i+\eta-d^z\eta)\sin(+c_i-\eta+d^z\eta)}
        {\sin(\frac{\eta}{2}+\xi)}]\\
& &\left. +\;\frac{2\sin(2\xi)\sin(c_i-d^z\eta)\sin(c_i+\eta-d^z\eta)
                            \sin(c_i+d^z\eta)\sin(c_i-\eta+d^z\eta)}
                {\sin(\frac{\eta}{2}-\xi)\sin(\frac{\eta}{2}+\xi)}
                                          \right \},\\
D_i&=&\frac{\sin(c_i-\frac{\eta}{2}-d^z\eta+\xi)}
           {2\sin(\frac{\eta}{2}-\xi)\sin(\frac{\eta}{2}+\xi)}\\ 
   & &     \times\left [(2d^2\sin^2\eta-\cos\eta)\sin(\frac{\eta}{2}-\xi)
             -\cos(2c_i)\sin(\frac{\eta}{2}+\xi)
             +\sin\eta\cos(\frac{5\eta}{2}+2d^z\eta-\xi)\right ],\\
E_i&=&\frac{\sin^3\eta\sin(c_i-\frac{3\eta}{2}-d^z\eta+\xi)
                          \sin(c_i-\frac{\eta}{2}-d^z\eta+\xi)}
           {2\cos\eta\sin(\frac{\eta}{2}-\xi)\sin(\frac{\eta}{2}+\xi)},\\
F_i&=&\frac{\sin(c_i-\frac{\eta}{2}-d^z\eta+\xi)}
{2\sin(\frac{\eta}{2}-\xi)\sin(\frac{\eta}{2}+\xi)}\\
   & &     \times\left [-(2d^2\sin^2\eta-\cos\eta)\sin(\frac{\eta}{2}+\xi)
             +\cos(2c_i)\sin(\frac{\eta}{2}-\xi)
             -\sin\eta\cos(\frac{\eta}{2}-2d^z\eta+\xi)\right ],
\end{eqnarray*}
where $i=a,b,$ and $\xi$ should be changed to $\xi^+$ when $i=b$. As in the spin-1/2
case, one can check that the Hamiltonian is
hermitician when we choose pure imaginary $c_i$. 
\\
\\
{\bf 3.4 $\;\;\;$ The Bethe ansatz for this model}
\\
\\
\par To construct the algebraic Bethe ansatz, we define the 
pseudo-vacuum state $\vert
0\rangle$ as
\begin{eqnarray}
      s_i^+\vert 0\rangle&=&d^+\vert 0\rangle=0,\;\;(i=1,\;\;2,\;\;\cdots,\;\;N) 
                                                         \nonumber\\
           d^z\vert 0\rangle&=&m\vert 0\rangle.
 \end{eqnarray}
 And as before, we write $T(u)$ as
\begin{equation}
T(u)=\left ( \begin{array}{lll}
{\cal A}_1(u)& {\cal B}_1(u)&{\cal B}_2(u)\\
{\cal C}_1(u)& {\cal A}_2(u)&{\cal B}_3(u)\\
{\cal C}_2(u)& {\cal C}_3(u)&{\cal A}_3(u) 
\end{array}\right ). \end{equation}
In order to simplify our calculation, we introduce the following transformations
\begin{eqnarray}
{\tilde {\cal A}}_2(u)&=&{\cal A}_2(u)-\frac{\sin(2\eta)}{\sin(2u+2\eta)}{\cal
A}_1(u)\\
{\tilde {\cal A}}_3(u)&=&{\cal A}_3(u)-\frac{\sin(2\eta)}{\sin(2u)}{\cal A}_2(u)
 -\frac{\sin\eta\sin(2\eta)}{\sin(2u+\eta)\sin(2u+2\eta)}{\cal A}_1(u)
\end{eqnarray}

It is easy to show 
$$ \begin{array}{lll}
{\cal C}_i(u)\vert 0\rangle=0,&{\cal B}_i(u)\vert 0\rangle\ne 0,&
                    (i=1,\;2,\;3),\\
{\cal A}_1(u)\vert 0\rangle=w_1\vert 0\rangle, &
{\tilde{\cal A}}_2(u)\vert 0\rangle=w_2\vert 0\rangle,&
{\tilde{\cal A}}_3(u)\vert 0\rangle=w_3\vert 0\rangle, 
\end{array}$$
with
\begin{eqnarray*}
w_1&=&\sin^{2N}(u+\eta)\sin^{2N}(u+2\eta)\\
   & &  \times \prod_{i=a,b}\prod_{r=\pm 1}
                   \sin(u+rc_i+m\eta)\sin(u+rc_i+\eta+m\eta),\\
w_2&=&\frac{\sin(2u)\sin(\xi-u-\frac{3\eta}{2})}
           {\sin(2u+2\eta)\sin(\xi+u+\frac{\eta}{2})}\sin^{2N} u\sin^{2N}(u+\eta)\\
& &           \times\prod_{i=a,b}\prod_{r=\pm 1}
           [\sin(u+rc_i)\sin(u+rc_i+\eta)+d^2\sin^2\eta-2\sin^2(m\eta)\cos\eta,\\
w_3&=&\frac{\sin(-\xi+u+\frac{\eta}{2})\sin(-\xi+u+\frac{3\eta}{2})\sin(\eta-2u)}
           {\sin(\xi+u+\frac{\eta}{2})\sin(-\xi-u+\frac{\eta}{2})\sin(\eta+2u)}
               \sin^{2N} u\sin^{2N}(u-\eta) \\ & &
            \times \prod_{i=a,b}\prod_{r=\pm 1}
                   \sin(u+rc_i-m\eta)\sin(u+rc_i+\eta-m\eta).
\end{eqnarray*}
From the reflection equation (5), we can find the following commutation relation
\begin{eqnarray}
{\cal A}_1(u){\cal B}_1(v)\vert 0\rangle
&=&\frac{\sin(u-v-2\eta)\sin(u+v)}{\sin(u+v+2\eta)\sin(u-v)}
                      {\cal B}_1(v){\cal A}_1(u)\vert 0\rangle \nonumber \\
&+&\frac{\sin(2v)\sin(2\eta)}{\sin(u-v)\sin(2v+2\eta)}
                      {\cal B}_1(u){\cal A}_1(v)\vert 0\rangle \nonumber \\
&-&\frac{\sin(2\eta)}{\sin(u+v+2\eta)}{\cal B}_1(u){\tilde{\cal A}}_2(v)\vert 0\rangle
\end{eqnarray}
\begin{eqnarray}
& &{\tilde{\cal A}}_2(u){\cal B}_1(v)\vert 0\rangle \nonumber \\
&=&\frac{\sin(u-v-2\eta)\sin(u-v+\eta)\sin(u+v)\sin(u+v+3\eta)}
        {\sin(u-v-\eta)\sin(u+v+\eta)\sin(u-v)\sin(u+v+2\eta)}
               {\cal B}_1(v){\tilde{\cal A}}_2(u)\vert 0\rangle \nonumber \\
&+&\frac{\sin(2v)\sin(2\eta)}{\sin(u-v-\eta)\sin(2v+2\eta)}
               {\cal B}_3(u){\cal A}_1(v)\vert 0\rangle 
 -\frac{\sin(2\eta)}{\sin(u+v+\eta)}
               {\cal B}_3(u){\tilde{\cal A}}_2(v)\vert 0\rangle \nonumber \\
&+&\frac{\sin(2\eta)[\sin(2\eta)\sin(u-v+\eta)-\sin(u+v+\eta)\sin(2u+2\eta)]}
        {\sin(u+v+\eta)\sin(2u+2\eta)\sin(u-v)}
               {\cal B}_1(u){\tilde{\cal A}}_2(v)\vert 0\rangle \\
&+&\frac{\sin(2\eta)\sin(2v)[\sin(2u+3\eta)\sin(u-v-2\eta)-\sin\eta\sin(u+v+2\eta)]}
        {\sin(u-v-\eta)\sin(u+v+2\eta)\sin(2u+2\eta)\sin(2v+2\eta)}
               {\cal B}_1(u){\cal A}_1(v)\vert 0\rangle, \nonumber \\
& &{\tilde{\cal A}}_3(u){\cal B}_1(v)\vert 0\rangle \nonumber \\
&=&\frac{\sin(u-v+\eta)\sin(u+v+3\eta)}{\sin(u-v-\eta)\sin(u+v+\eta)}
               {\cal B}_1(v){\tilde{\cal A}}_3(u)\vert 0\rangle  \nonumber \\
&-&\frac{\sin(2\eta)\sin(2v)\sin(2u+2\eta)}{\sin(2u)\sin(u-v-\eta)}
               {\cal B}_3(u){\tilde{\cal A}}_2(v)\vert 0\rangle  \nonumber \\
&+&\frac{\sin(2\eta)\sin(2v)\sin(2u+2\eta)}
        {\sin(2u)\sin(2v+2\eta)\sin(u+v+\eta)}
               {\cal B}_3(u){\cal A}_1(v)\vert 0\rangle  \nonumber \\
&+&\frac{\sin^2(2\eta)\sin(2u+2\eta)}
        {\sin(2u)\sin(2u+\eta)\sin(u-v-\eta)}
               {\cal B}_1(u){\tilde{\cal A}}_2(v)\vert 0\rangle  \nonumber \\
&-&\frac{\sin^2(2\eta)\sin(2v)\sin(2u+2\eta)}
        {\sin(2u)\sin(2u+\eta)\sin(2v+2\eta)\sin(u+v+2\eta)}
               {\cal B}_1(u){\cal A}_1(v)\vert 0\rangle, 
\end{eqnarray} 

 From the reflection equation (5), we can constructed a two-particle
excited state as follows,
   \begin{eqnarray}
   \vert v_1,v_2\rangle&\equiv& \left\{{\cal B}_1(v_1){\cal B}_1(v_2)
          +\frac{\sin(2\eta)}{\sin v_1+\sin v_2+\eta}
                       {\cal B}_2(v_1){\cal A}_2(v_2) \right.\nonumber\\
& &\;\;-\;\left.\frac{\sin(2\eta)\sin(v_1+v_2-\eta)}
              {\sin(v_1-v_2-\eta)\sin(v_1+v_2+\eta)}
                       {\cal B}_2(v_1){\cal A}(v_2)\right\}\vert 0 \rangle
  \end{eqnarray}
   which is symmetric in $v_1,v_2$ up to a whole factor.
Applying the transfer matrix on the two-particle excited state, we have found
a lot of ``unwanted terms". They must cancel each other to ensure the
above state to be eigenstate. However, we can't check it directly
because the calculation is much more complicated than we  expected. 
Here  we
assume that they vanish. To get the Bethe ansatz equations, we have to 
use the 
functional Bethe ansatz method  which was first proposed for 
the Ising model\cite{B}.
   Tarasov $\sl{et\;\; al}$ argued that this method  can be
generalized for n-particle
excited state[29-31]. Then from equation (8), the 
eigenvalue of the
transfer matrix for n-particle excited states is as follows
\begin{eqnarray} 
& &{\tilde t}(u)\vert v_1,v_2,\cdots,v_n\rangle=(
        k_1^+{\cal A}_1(u)+k_2^+{\cal A}_2(u)+k_3^+{\cal A}_3(u))\vert
                                      v_1,v_2,\cdots,v_n\rangle\nonumber\\
&=&(w_1^+{\cal A}_1(u)+w_2^+{\tilde{\cal A}}_2(u)+w_3^
        +{\tilde{\cal A}}_3(u))\vert v_1,v_2,\cdots,v_n\rangle \nonumber\\
&=&w_1^+w_1\prod_{i=1}^n\frac{\sin(u-v_i-2\eta)\sin(u+v_i)}{\sin(u+v_i+2\eta)\sin(u-v_i)}
                              \vert v_1,v_2,\cdots,v_n\rangle \nonumber\\
&+&w_2^+w_2\prod_{i=1}^n\frac{\sin(u-v_i-2\eta)\sin(u-v_i+\eta)\sin(u+v_i)\sin(u+v_i+3\eta)}
{\sin(u-v_i-\eta)\sin(u+v_i+\eta)\sin(u-v_i)\sin(u+v_i+2\eta)}       
                               \vert v_1,v_2,\cdots,v_n\rangle \nonumber\\
&+&w_3^+w_3\prod_{i=1}^n\frac{\sin(u-v_i+\eta)\sin(u+v_i+3\eta)}{\sin(u-v_i-\eta)\sin(u+v_i+\eta)}
                               \vert v_1,v_2,\cdots,v_n\rangle  
\end{eqnarray}
where 
\begin{eqnarray*}
w_1^+(u)&=&\frac{\sin(2u+3\eta)\sin(u-\xi^+-\frac{\eta}{2})}
                {\sin(2u+\eta)\sin(u-\xi^++\frac{3\eta}{2})},\\
w_2^+(u)&=&\frac{\sin(\xi^++u+\frac{3\eta}{2})\sin(u-\xi^+-\frac{\eta}{2})\sin(2u+2\eta)}
{\sin(-\xi^++u+\frac{3\eta}{2})\sin(-u+\xi^+-\frac{\eta}{2})\sin(2u)},\\
w_3^+(u)&=&\frac{\sin(\xi^++u+\frac{3\eta}{2})\sin(-u-\xi^+-\frac{\eta}{2})}
{\sin(-\xi^++u+\frac{3\eta}{2})\sin(-u+\xi^+-\frac{\eta}{2})},
\end{eqnarray*}
 and the free parameters $v_i$,$i$=1,2,$\cdots$,$n$ obey  the Bethe ansatz
equation
\begin{eqnarray}
& 
&\frac{\sin(\xi^+-v_j+\frac{\eta}{2})\sin(\xi+v_j-\frac{\eta}{2})\sin^{2N}(v_j+\eta)}
{\sin(\xi^++v_j+\frac{\eta}{2})\sin(\xi-v_j-\frac{\eta}{2})\sin^{2N}(v_j-\eta)}
\prod_{k=a,b}\prod_{r=\pm 1}
\frac{\sin(v_j+rc_k+m\eta)}
     {\sin(v_j+rc_k-m\eta)}\nonumber\\
&=&\prod_{i\ne j}\frac{\sin(v_j-v_i+\eta)\sin(v_j+v_i+\eta)}
                      {\sin(v_j-v_i-\eta)\sin(v_j+v_i-\eta)},
   \ \ \ \ \ \ \ \ \ \  \;j=1,\;2,\;\cdots,\;n.
\end{eqnarray} 

This Bethe ansatz equation can also be derived by the means of the fusion
method, we have checked that they  agree with each other\cite{YB,Mez}.
 With this method of Bethe ansatz, we cannot get complete eigenstates. However,  it
is a powerful tool to obtain the eigenvalues and Bethe ansatz equation for models
which cannot be obtained with fusion method\cite{F}. 
\\
\\
{\bf 4. $\;\;\;$  Discussion}
\\
\\
In this paper, we  studied the integrability of the 
spin-1/2 and spin-1 XXZ open Heisenberg chains with
boundary  impurities. These models are relevant to the Kondo problem 
in a Luttinger liquid. By using the algebraic Bethe
ansatz and its extension, we have 
obtained the eigenvalues of the Hamiltonians and 
the Bethe ansatz equations. When we let $\vec d=0$ in this paper, 
one can easily check that the  $L_i(u)\;(i=a,\;b)$ in formulas (10)
and(40) will be
identity so the present Hamiltonians and Bethe ansatz equations 
can be reduced to the usual ones respectively. 
This procedure can be generalized 
to the general Heisenberg chain. 
It is worthy to point out that  the Bethe ansatz equations and eigenvalues
of transfer matrices  for the spin-1/2 and spin-1
XXX chains coupled with arbitrary spin impurities can be obtained by
rescaling the
spectral parameters $v_j$ by $v_j\times\eta$ and taking the limit $ \eta
\rightarrow 0$ in Bethe ansatz equations and the eigenvalues of the
transfer matrices.    
With similar methods of Ref.\cite{y,bab,bab1},
the results of the present 
paper can also be used to calculate the boundary
susceptibility, the contribution of the impurities to the specific heat
and Kondo temperature, which can describe the effect of impurities to the 
system. We will study them in another paper\cite{zsy1}.

 \end{document}